\documentclass[aps,prl,twocolumn,superscriptaddress,showpacs,showkeys,amsfonts,amssymb,amsmath]{revtex4}
\usepackage{graphicx}
\usepackage{color}
\usepackage{ulem}
\usepackage[colorlinks=true,linkcolor=blue,citecolor=blue,urlcolor=blue]{hyperref}
\begin{document}

\title{Magneto- to electro-active transmutation of spin waves in ErMnO$_3$}

\author{L. Chaix}
\email{chaixl@ill.fr}
\affiliation{Institut Laue-Langevin, 6 rue Jules Horowitz,38042 Grenoble, France}
\affiliation{Universit\'e Grenoble Alpes, Institut N\'eel, 38042 Grenoble, France}
\affiliation{CNRS, Institut N\'eel, 38042 Grenoble, France}
\author{S. de Brion}
\email{sophie.debrion@neel.cnrs.fr}
\affiliation{Universit\'e Grenoble Alpes, Institut N\'eel, 38042 Grenoble, France}
\affiliation{CNRS, Institut N\'eel, 38042 Grenoble, France}
\author{S. Petit}
\affiliation{CEA, Centre de Saclay, /DSM/IRAMIS/ Laboratoire L\'eon Brillouin, 91191 Gif-sur-Yvette,France}
\author{R. Ballou}
\affiliation{Universit\'e Grenoble Alpes, Institut N\'eel, 38042 Grenoble, France}
\affiliation{CNRS, Institut N\'eel, 38042 Grenoble, France}
\author{L.-P. Regnault}
\affiliation{SPSMS-MDN, UMR-E CEA/UJF-Grenoble, INAC, 38054 Grenoble, France}
\author{J. Ollivier}
\affiliation{Institut Laue-Langevin, 6 rue Jules Horowitz,38042 Grenoble, France}
\author{J.-B. Brubach}
\affiliation{Synchrotron SOLEIL, L'Orme des Merisiers Saint-Aubin, 91192 Gif-sur-Yvette, France}
\author{P. Roy}
\affiliation{Synchrotron SOLEIL, L'Orme des Merisiers Saint-Aubin, 91192 Gif-sur-Yvette, France}
\author{J. Debray}
\affiliation{Universit\'e Grenoble Alpes, Institut N\'eel, 38042 Grenoble, France}
\affiliation{CNRS, Institut N\'eel, 38042 Grenoble, France}
\author{P. Lejay}
\affiliation{Universit\'e Grenoble Alpes, Institut N\'eel, 38042 Grenoble, France}
\affiliation{CNRS, Institut N\'eel, 38042 Grenoble, France}
\author{A. Cano}
\affiliation{European Synchrotron Radiation Facility, 6 rue Jules Horowitz, BP 220, 38043 Grenoble, France}
\affiliation{CNRS, Univ. Bordeaux, ICMCB, UPR 9048, F-33600 Pessac, France}
\author{E. Ressouche}
\affiliation{SPSMS-MDN, UMR-E CEA/UJF-Grenoble, INAC, 38054 Grenoble, France}
\author{V. Simonet}
\affiliation{Universit\'e Grenoble Alpes, Institut N\'eel, 38042 Grenoble, France}
\affiliation{CNRS, Institut N\'eel, 38042 Grenoble, France}

\date{\today}

\begin{abstract}
{The low energy dynamical properties of the multiferroic hexagonal perovskite ErMnO$_3$ have been studied by inelastic neutron scattering as well as  terahertz and far infrared spectroscopies on synchrotron source. From these complementary techniques, we have determined the magnon and crystal field spectra and identified a zone center magnon only excitable by the electric field of an electromagnetic wave. Using comparison with the isostructural YMnO$_3$ compound and crystal field calculations, we propose that this dynamical magnetoelectric process is due to the hybridization of a magnon with an electro-active crystal field transition.}
\end{abstract}


\maketitle


The term magnetoelectric (ME) makes reference to a variety of phenomena in which electric dipoles and magnetic moments are mutually linked \cite{Fiebig2002a}. ME processes attract a considerable research interest, largely driven by their potential use in future information technologies \cite{Fiebig2002a,Bibes} and, more recently, by their interpretation in terms of exotic magnetic and ME monopoles \cite{monopoles}. A ME process that is particularly striking is the electric-charge dressing of spin-waves that gives rise to electrically active magnons (or electromagnons) \cite{electromagnons}.This dynamical ME effect has been most clearly demonstrated in multiferroic materials such as $R$MnO$_3$ \cite{em-RMnO3}, $R$Mn$_2$O$_5$ \cite{em-RMn2O5}, CuFeO$_2$ \cite{em-CuFeO}, and BiFeO$_3$ \cite{em-BiFeO}. The ME dual of electromagnons, that is magneto-active excitations of the lattice, have also been recently reported \cite{Chaix2013}. This type of hybrid excitations enable additional optical functionalities such as directional light switching and quadrochroism \cite{optics} that open new perspectives in photonics and magnonics \cite{Kruglyak2010}.

 \begin{figure}
\resizebox{7.6cm}{!}{ \includegraphics{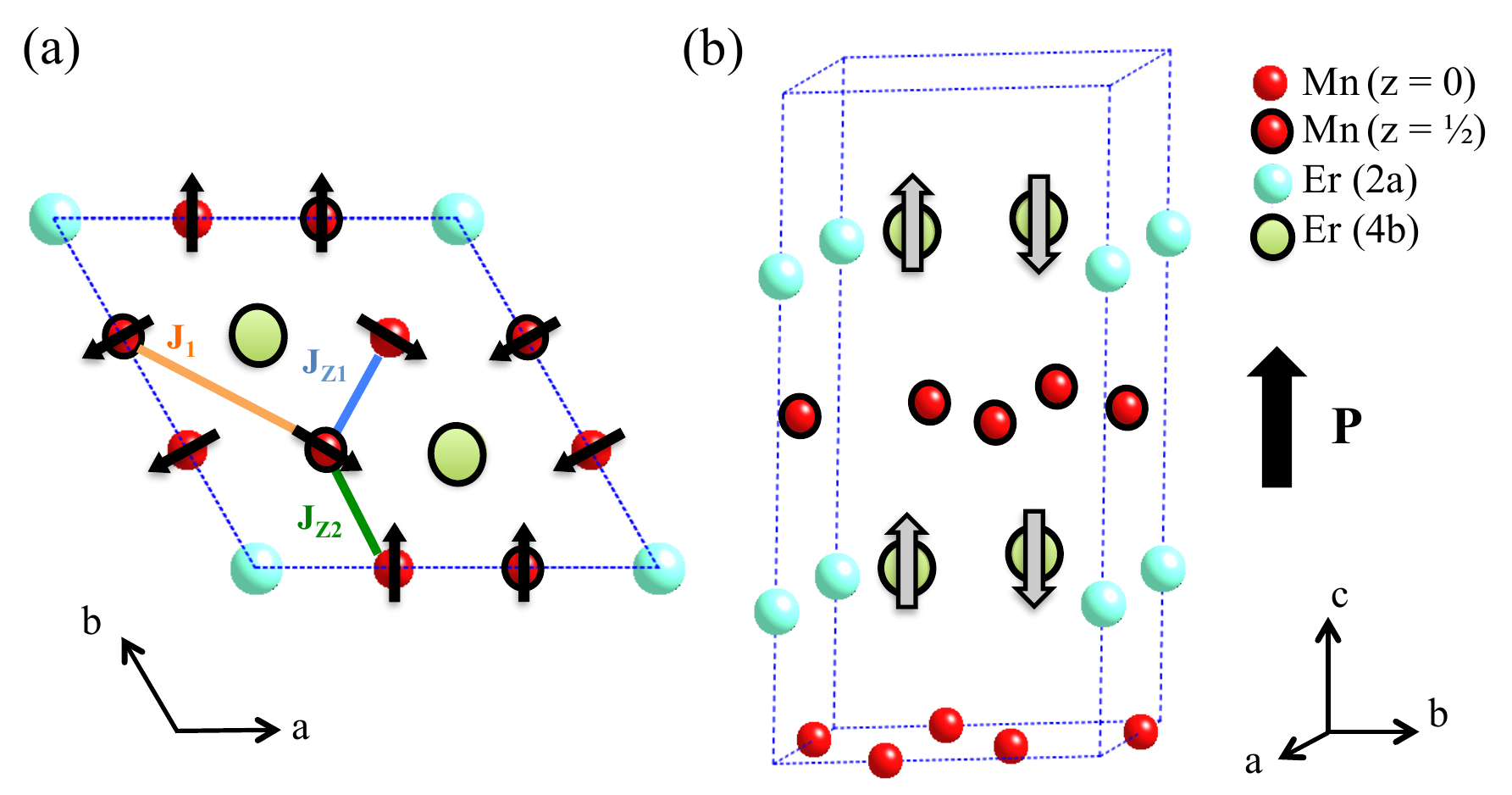}}
 \caption{Schematic illustration of ErMnO$_3$ crystallographic and magnetic structure in the temperature range $T'_N<T<T_N$ where Mn magnetic moments are ordered at 120$^\circ$ (a) and Er moments on the $4b$ site only are polarized in an antiferromagnetic arrangement (b). This multiferroic phase presents an electric polarisation $\bf{P}$ along the $\bf{c}$-axis. The exchange interactions involved in the Mn magnetic order are also shown.}
 \label{fig1}
 \end{figure}

Several microscopic mechanisms have been identified behind the ME character of these hybrid excitations. In essence, they all trace back to the specific couplings between spins and the deformable lattice that can produce ferroelectricity. In orthorhombic $R$MnO$_3$, for example, both the so-called inverse Dzyaloshinksii-Moriya and Heisenberg exchange mechanisms contribute to the static polarization \cite{P-RMnO3} and also generate different electromagnons \cite{em-RMnO3}. In hexagonal YMnO$_3$, a phonon-magnon anticrossing has been observed revealing that lattice and spins are dynamically coupled in this system too \cite{Petit2007,Paihles2009}.
However, the ME nature of this feature was not been proven and, to the best of our knowledge, no electromagnon has been reported so far in the other members of this family of prominent multiferroics \cite{Liu2012,Standard2012,Toulouse}. In this Letter, we report on the presence of a new type of ME excitation in the hexagonal ErMnO$_3$ multiferroic compound. By means of complementary spectroscopic tools and the comparison with hexagonal
YMnO$_3$, we show evidence of
the transmutation of a regular magnon to an electro-active excitation in ErMnO$_3$. We ascribe this effect to a distinct hybridization mechanism between Mn$^{3+}$ spin-waves and crystal field (CF) excitations of the Er$^{3+}$ rare earth.

 \begin{figure*}
\resizebox{15cm}{!}{ \includegraphics{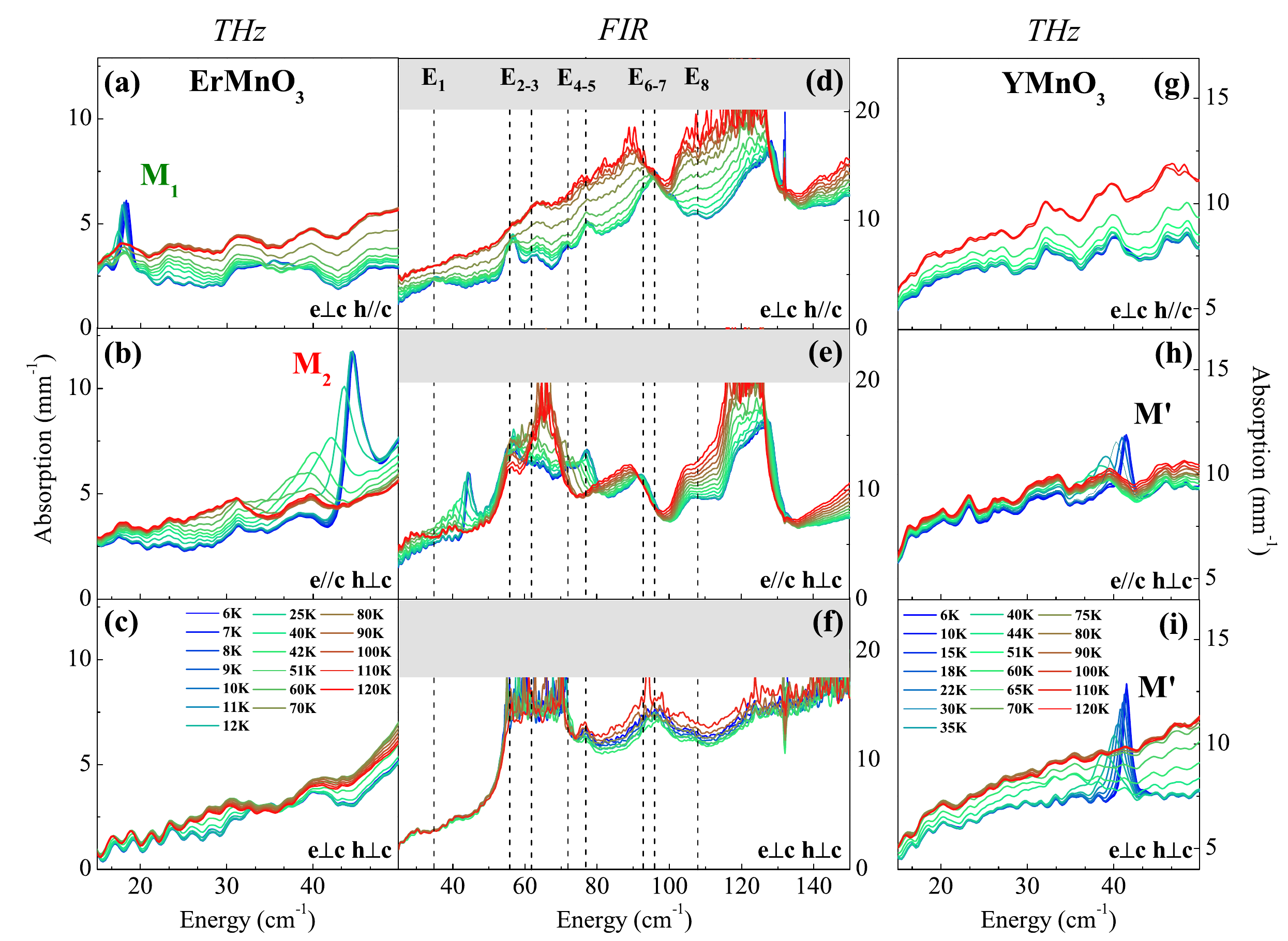}}
 \caption{THz (a,b,c) and FIR (d,e,f) absorption spectra of ErMnO$_3$ for the 3 different orientations of  the electromagnetic wave {\bf e}, {\bf h} fields with respect to the crystal $\bf{c}$-axis,  in the temperatures range 6 K - 120 K. The grey areas indicate regions where the absorption saturates. The various excitation modes discussed in the text are pointed out. THz (g,h,i) absorption spectra of YMnO$_3$ measured in the same conditions as ErMnO$_3$.}
 \label{fig2}
 \end{figure*}

In ErMnO$_3$, the ferroelectric state occurs below T$_c$ = 833 K with a spontaneous polarization along the $\bf{c}$-axis (polar space group $P6_3cm$). Mn$^{3+}$ ions, located on the $6c$ Wyckoff sites, form a triangular lattice in the ($\bf{a}$,  $\bf{b}$) planes. A complex magnetic order with a 120$^\circ$ arrangement in the ($\bf{a}$,  $\bf{b}$) plane ($\bf{k} = 0$ propagation vector) takes place at T$_N$ = 79 K, primarily produced by the Mn$^{3+}$ magnetic moments ($\Gamma_4$ irreducible representation \cite{Meier2012,Chaix}). On the other hand, the Er$^{3+}$ ions occupy two distinct Wyckoff sites, $4b$ and $2a$ \cite{Fiebig2002b}. Below T$_N$, the $4b$ magnetic moments are orientated antiferromagnetically along the $\bf{c}$-axis, polarized in the molecular field of the Mn magnetic moments \cite{Fabreges2009} (See Fig.\ref{fig1}). Below T'$_N\approx$ 10 K, the $2a$ magnetic moments order ferromagnetically along the $\bf{c}$-axis and produce a spin reorientation at the $4b$ rare-earth and Mn sites around 5 and 2 K respectively ($\Gamma_2$ irreducible representation \cite{Meier2012,Chaix}).
Besides its multiferroic properties, ErMnO$_3$ is particularly interesting because of the strong interplay between the Mn$^{3+}$ ($3d^4$) magnetism and the Er$^{3+}$ ($4f^{11}$) spins \cite{Meier2012,Chaix}.


We probed the dynamical properties of ErMnO$_3$ by inelastic neutron scattering and THz/FIR spectroscopies. Additionally, in order to confirm the strong influence of the Er magnetism on the dynamical properties of ErMnO$_3$, we also measured the THz response of YMnO$_3$ where Er is replaced by non-magnetic Y. THz/FIR measurements were performed on the AILES beamline at SOLEIL Synchrotron using a Bruker IFS125 interferometer \cite{Roy2006}. This technique allows to probe magnetic and electric excitations, their characteristic energies as well as their selection rules as regards the electric ({\bf e}) and magnetic ({\bf h}) components of the polarized electromagnetic wave. For this purpose, two disks (thickness $\approx$~500$\mu m$, surface $\approx$~12~$mm^{2}$) were cut in the same single crystal grown by floating zone method in an image furnace, with the $\bf{c}$-axis in the plane and out of the plane perpendicular to the THz/FIR propagation vector. Using the same procedure as in \cite{Chaix2013}, absorption spectra were obtained in the temperature 6 - 120 K with three different geometries:
i) {\bf e$\perp$c}~~{\bf h$\parallel$c}, ii) {\bf e$\parallel$c}~~{\bf h$\perp$c} and iii) {\bf e$\perp$c}~~{\bf h$\perp$c}. The THz (10 - 50~cm$^{-1}$) and FIR (20 - 200~cm$^{-1}$) range were explored at a resolution of 0.5~cm$^{-1}$ using the same 6~$\mu m$ thick silicon-mylar beam splitter and two different bolometers.

Whereas these experiments mainly probe the Brillouin zone center, inelastic neutron scattering is sensitive to a large volume of the reciprocal space. A first experiment using unpolarized neutrons was performed at the Institut Laue-Langevin (ILL) high-flux reactor in Grenoble on the time-of-flight spectrometer IN5 with an incident wavelength of 2 \AA\ for an energy resolution of 0.8 meV (7 cm$^{-1}$). To discriminate between the magnetic and nuclear contributions, we also used polarized neutrons and longitudinal polarization analysis on the CRG-CEA triple-axis spectrometer IN22 installed at the ILL \cite{Chaix}. The final energy was kept fixed at 14.7 meV yielding an energy resolution of 1 meV (8 cm$^{-1}$). In both experiments, performed at 1.5 K and 20 K respectively, the ErMnO$_3$ single-crystal was oriented with the $\bf{a}$-axis vertical in order to survey the ($\bf{b}$*, $\bf{c}$*) scattering plane.

Fig.\ref{fig2} gives an overview of the absorption spectra of ErMnO$_3$ (THz and FIR) and YMnO$_3$ (THz) and of their temperature dependence.
In ErMnO$_3$, two excitations $\bf{M_1}$ and $\bf{M_2}$ are clearly seen at low temperatures in the THz range (centered at 18.5 cm$^{-1}$ and 44.6 cm$^{-1}$ respectively at 6 K). The magnetic origin of these excitations is suggested by their softening and eventually disappearance at T$_N$. The first mode $\bf{M_1}$ is excited by the magnetic field {\bf h$\parallel$c}, as expected for a magnon.
The second excitation $\bf{M_2}$, in contrast, disappears when the electric field changes from {\bf e$\parallel$c} to {\bf e$\perp$c}, even if the magnetic field is kept {\bf h$\perp$c}. This implies that, unexpectedly, $\bf{M_2}$ is excited by the electric field of the THz wave.
In YMnO$_3$ a similar excitation $\bf{M'}$ is also seen at approximately the same energy (41.5 cm$^{-1}$ at 6 K). $\bf{M'}$, however, remains visible for {\bf h$\perp$c} irrespective of the orientation ({\bf e$\perp$c} or {\bf e$\parallel$c}) of the electric field as already reported \cite{Kadlec2011}. In fact, the comparison with the magnon dispersion curves analysed in \cite{Petit2007,Toulouse} confirms that $\bf{M'}$ is a magnon involving spin components perpendicular to the $\bf{c}$-axis in agreement with its {\bf h$\perp$c} excitation rule. This highlights the peculiarity of the ErMnO$_3$ $\bf{M_2}$ excitation which is only excited by the electric field of the THz wave.

In addition to these modes, FIR ErMnO$_3$ spectra \cite{footnote2} reveal additional low-energy excitations persisting above T$_N$. Their number and intensity depend on the temperature and on the orientation of the electric/magnetic fields (see Fig.\ref{fig2}). At low temperature, there is a first weak excitation at 35 cm$^{-1}$ ({\bf E$_1$}). Then, there is a set of 4 excitations at 56, 62, 72 and 77 cm$^{-1}$ ({\bf E$_2$}, {\bf E$_3$}, {\bf E$_4$} and {\bf E$_5$}) that merge into three excitations above T$_N$. Note that these signals are enhanced for {\bf e$\parallel$c}~~{\bf h$\perp$c} in the sample with in plane $\bf{c}$-axis. Next, the two close excitations at 93 and 96 cm$^{-1}$ ({\bf E$_6$} and {\bf E$_7$}) also merge into a single one above T$_N$. Another excitation is visible around 108 cm$^{-1}$ ({\bf E$_8$}). We note that this identification is consistent with the modes reported in reference \onlinecite{Standard2012} from FIR spectroscopy, except for the weaker modes $\bf E_1$ and $\bf E_8$ \cite{footnote1}. Finally, around 125 cm$^{-1}$ a larger phononic band is observed in agreement with Ab initio calculation \cite{Iliev1997} and Raman spectroscopy \cite{Vermette2008}.

\begin{figure}
\resizebox{8cm}{!}{\includegraphics{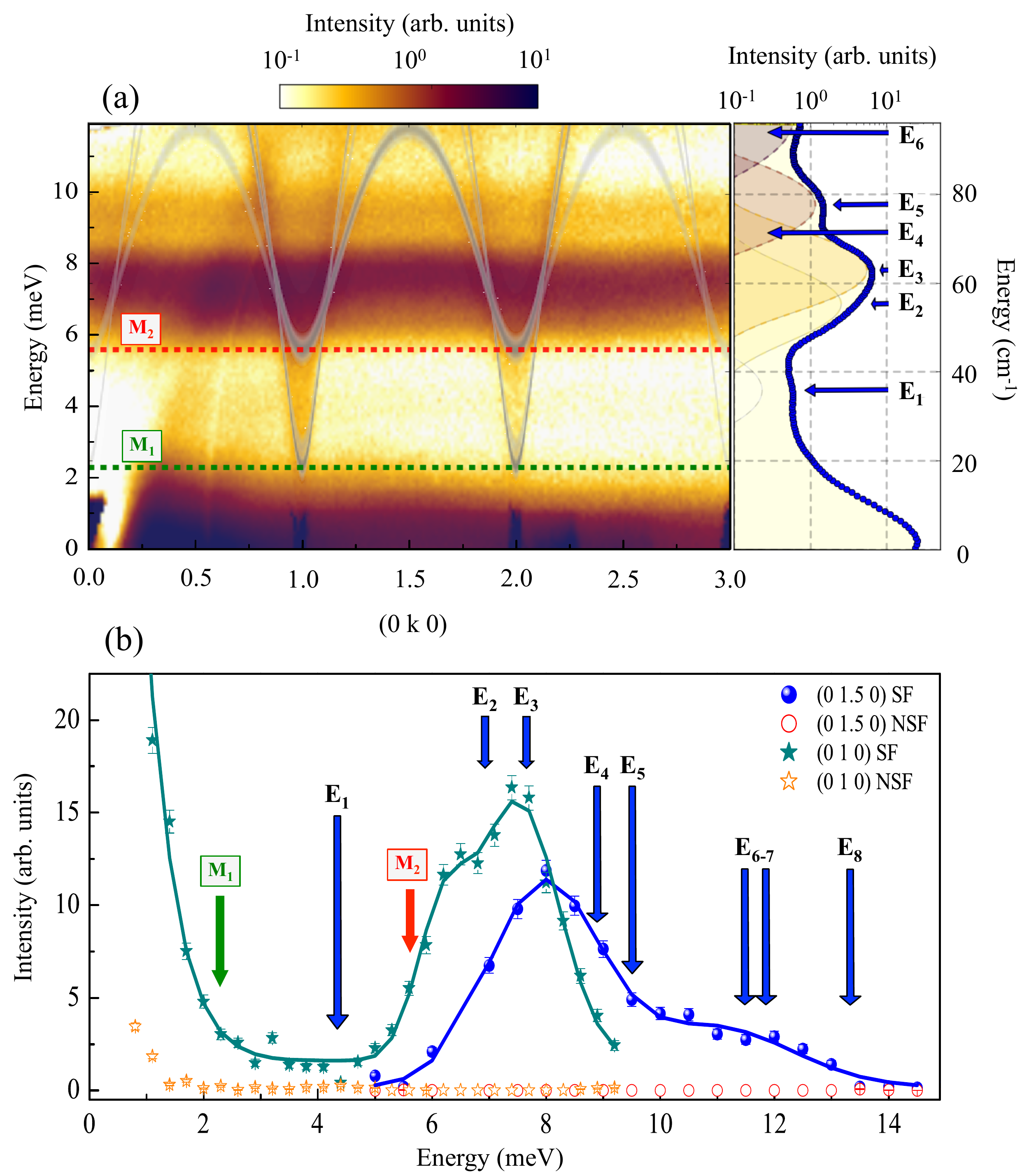}}
\caption{(a)Left: IN5 inelastic neutron signal spectrum recorded at 1.5 K along (0 $k$ 0), the data were cut and integrated in a reciprocal space part, from $\ell$ = -2 to +2 along the $\bf{c^*}$ direction. The calculated Mn spin-waves dispersion (identical for $\Gamma_2$ or $\Gamma_4$) are reported (continuous lines). The corresponding Hamiltonian involves three exchange interactions, one in-plane first neighbor $J_1=2.65$ meV, and two out-of-plane interactions $J_{z1}$ and $J_{z2}$ with $J_{z1} - J_{z2} = -0.007$ meV (see Fig.\ref{fig1}), as well as two anisotropy terms at $D = 0.39$ meV and $h = 0.055$ meV producing the energy gaps for the spin components in and perpendicular to the ($\bf{a}$, $\bf{b}$) plane respectively. Right: corresponding energy map obtained by powder-average  of all  single crystal spectra;  all scattering angles were grouped assuming nearly non dispersive CF excitations (b) IN22 spin flip and non spin flip neutron scattering signals recorded at 20 K at the (0 1 0) zone center and (0 1.5 0) zone boundary. The magnon modes {\bf M$_i$} and CF transitions {\bf E$_i$} observed in THz and FIR spectroscopy are also shown.}
\label{fig3}
\end{figure}


The observed ErMnO$_3$ excitations can be either of magnonic origin, either due to transitions between CF levels of the two Er ion types, or from phonons. Inelastic neutron scattering allows to further identify their origin. Fig.\ref{fig3}a presents IN5 measurements at 1.5 K along the (0 $k$ 0) direction. A dispersive signal emerges from the two (0 1 0) and (0 2 0) zone centers and corresponds to spin-waves associated to the Mn$^{3+}$ magnetic order. This signal was well characterized combining IN5 and IN22 measurements and can be reproduced by spin-wave calculations in the linear approximation of the Holstein-Primakoff formalism \cite{HolsteinPrimakoff1940}, using the same Hamiltonian as in reference \onlinecite{Fabreges2009} (see the continuous lines in Fig.\ref{fig3}a and reference \onlinecite{Chaix}). The neutron signal is broadened by convolution to the limited experimental resolution in these instrumental configurations.

In the (0 $k$ 0) zone centers, the spin wave of the lowest branch at 18 cm$^{-1}$ is gapped due to the coupling between the Mn and the anisotropic Er ions and correlates  spin components along the $\bf{c}$-axis at the (0 0 0) zone center. It thus agrees with the $\bf {M_1}$ THz mode excited by {\bf h$\parallel$c}. Note that this gap is reduced in YMnO$_3$ where the lowest branch lies out of the range of the THz measurements. In ErMnO$_3$, the higher energy branch is calculated at 45 cm$^{-1}$, which corresponds well to the $\bf{M_{2}}$ excitation and is similar to the one observed in YMnO$_3$ at 41.5 cm$^{-1}$. The calculations show that it involves spin components perpendicular to the $\bf{c}$-axis and should therefore be observed for {\bf h$\perp$c}, as it is the case in YMnO$_3$. However it is not observed experimentally in the configuration {\bf e$\perp$c}~~{\bf h$\perp$c} in ErMnO$_3$. The $\bf {M_2}$ excitation is therefore sensitive to the electric component of the incident wave, although it has a magnetic origin since it coincides with the energy of a magnon and disappears above T$_N$.

The Mn spin-waves dispersive signal goes through flatter broad excitations. A closer look using the integrated intensity in the (0 k 0) direction (left panel in Fig.\ref{fig3}a) allows to distinguish 5 excitation modes corresponding to those observed in THz and FIR spectroscopy. Moreover, as shown from measurements with polarized neutron at 20 K (Fig.\ref{fig3}b), these modes have essentially a s spin-flip character i.e. associated with magnetic scattering, and are thus attributed to CF level transitions. They correspond to {\bf E$_1$} (35 cm$^{-1}$), to {\bf E$_2$} and {\bf E$_3$} (60 cm$^{-1}$), to {\bf E$_4$} and {\bf E$_5$} (75 cm$^{-1}$), to {\bf E$_6$} and {\bf E$_7$} (97 cm$^{-1}$) and to {\bf E$_8$}, all observed with a better energy resolution in FIR measurements. Last, the neutron scattering experiments also reveal the strong coupling between the Mn$^{3+}$ and Er$^{3+}$ $4b$ magnetism at the dynamical level with, in particular, a noticeable modulation of the 60~cm$^{-1}$ broad band at the crossing points with the Mn$^{3+}$ spin-waves. Another indication of this coupling is the Zeeman splitting of CF levels seen in FIR spectroscopy below T$_N$, and also reported in reference \onlinecite{Standard2012}, produced by the molecular field arising from the Mn magnetic order: the four {\bf E$_2$} to {\bf E$_5$} excitations and the levels around 98 cm$^{-1}$ ({\bf E$_6$} and {\bf E$_7$}) probably results from three excitations and from a single one respectively through cooling below T$_N$.

\begin{figure}
\resizebox{7cm}{!}{\includegraphics{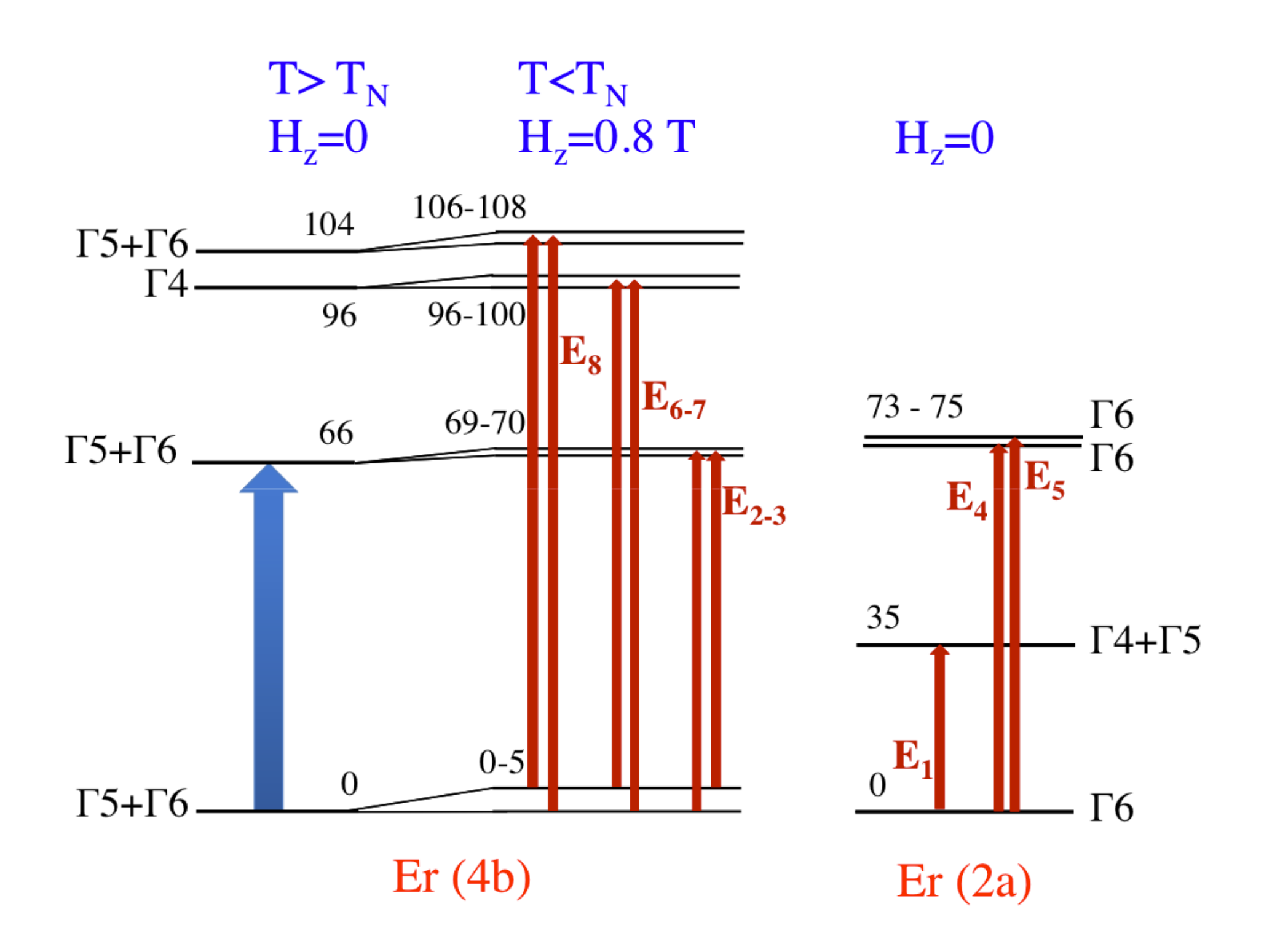}}
\caption{Crystal field calculation from a point charge model for the two $4b$ and $2a$ Er$^{3+}$ sites. The CF levels are labelled according to the irreducible representations of the double point groups associated with each Er sites. The splitting produced by a magnetic field of 0.8 T mimics the effect of the Mn molecular field on Er($4b$) CF levels, which occurs below T$_N$. The CF levels scheme consists in 9 doublets for each sites. The modes above 200 cm$^{-1}$ are not shown. Note that the energy of the levels is very sensitive to the O$^{2-}$ coordinates, which were obtained from single-crystal neutron diffraction at 10 K \cite{Chaix}. The observed THz and FIR transitions {\bf E$_i$} are shown (red arrows) as well as the possible excitation involved in the hybridization with a Mn magnon (blue arrow).}
\label{fig4}
\end{figure}


From the above experimental determination of magnons and CF transitions scheme in ErMnO$_3$, we can now discuss the nature of the remarkable electro-active excitation $\bf{M_2}$. The equivalent excitation behaves as a standard magneto-active magnon in YMnO$_3$. This difference suggests that the electrical activity of the $\bf{M_2}$ excitation in ErMnO$_3$ is associated to the magnetic coupling between the magnetic rare earth and the Mn ions. It is known that a transition between CF levels of a rare-earth ion with a non-centrosymmetric point group can be either electro-active, or magneto-active, or both \cite{Mukhin1991}.
 Then, by hybridization, a magnetoelectric CF excitation can transfer its electro-activity to a magnon. This mechanism was already proposed in HoMn$_2$O$_5$ and Tb$_3$Fe$_5$O$_{12}$ from FIR measurements \cite{Sirenko2008,Kang2010}. In order to test the validity of this scenario in ErMnO$_3$, the Er$^{3+}$ (J=15/2) CF levels were calculated using a point charge model \cite{Stevens1952,Hutchings1965}, taking into account the 7 O$^{2-}$ closer neighbors of the two Er sites, with screening factors \cite{Newman2000}, fitted to recover the experimentally identified CF levels. The influence of a magnetic field was also computed to account for the splitting of the CF levels due to the coupling of the Er $4b$ with the Mn. The results are reported in Fig.\ref{fig4} and show that, within this simple model, a qualitative agreement is obtained with the experiment as regards the position and number of modes. Moreover, a symmetry analysis reveals that the two Er$^{3+}$ sites can have CF level transitions that are both electro- and magneto-active. Specifically, transitions between levels belonging to the same (different) irreducible representation of the point group symmetry can be activated by {\bf e$\parallel$c} and {\bf h$\parallel$c} ({\bf e$\perp$c} and {\bf h$\perp$c}). In view of this, the absence of the $\bf{M_2}$ excitation in the THz {\bf e$\perp$c} and {\bf h$\perp$c} configuration suggests that the hybridization occurs between the magnetically coupled Mn and Er $4b$ through a CF transition allowed for {\bf e$\parallel$c}. A possible transition is the one between the ground state level and the first excited level of Er($4b$), which is split below T$_N$ \cite{footnote3}.


In conclusion, we have observed the complete loss of the magnetic character of a magnon in ErMnO$_3$ transmuted to an electro-active excitation. We attribute this ME dynamical process to the hybridization between a CF level transition of the Er magnetic rare earth and a Mn magnon. This mechanism may be very general to other rare-earth based multiferroics. This exemplifies the richness of the emerging field of ME excitations and suggests new possibilities to manipulate these excitations, as for instance through the action of magnetic/electric static fields, that may dehybridized the CF transition and the magnon.

\acknowledgments This work was financially supported by the ANR-13-BS04-0013-01. We thanks L. Pinsard-Godart for providing the YMnO$_3$ crystals and X. Frabr\`eges for helpful discussions on the magnetic order and associated spin waves in hexagonal manganites.


\begin{thebibliography}
\bibitem{footnote3} The fact that there is no signature of this hybridization in the {\bf h//c} channel may be due to the different strength of the electroactive versus magnetoactive processes.
\end{thebibliography}
\end{document}